\documentclass[twocolumn]{aastex631}

\received{August 28, 2022}
\revised{October 19, 2022}
\accepted{November 1, 2022}

\shorttitle{Diverse Molecular Gas Properties in FRB Hosts}
\shortauthors{Hatsukade et al.}

\begin{document}

\title{Diverse Properties of Molecular Gas in the Host Galaxies of Fast Radio Bursts}

\author[0000-0001-6469-8725]{Bunyo Hatsukade}
\affiliation{Institute of Astronomy, Graduate School of Science, The University of Tokyo, 2-21-1 Osawa, Mitaka, Tokyo 181-0015, Japan}
\email{hatsukade@ioa.s.u-tokyo.ac.jp}

\author[0000-0001-7228-1428]{Tetsuya Hashimoto}
\affiliation{Department of Physics, National Chung Hsing University, No. 145, Xingda Rd., South Dist., Taichung, 40227, Taiwan (R.O.C.)}

\author[0000-0001-5322-5076]{Yuu Niino}
\affiliation{Research Center for the Early Universe, The University of Tokyo, 7-3-1 Hongo, Bunkyo, Tokyo 113-0033, Japan}
\affiliation{Institute of Astronomy, Graduate School of Science, The University of Tokyo, 2-21-1 Osawa, Mitaka, Tokyo 181-0015, Japan}

\author[0000-0002-0944-5634]{Tzu-Yin Hsu}
\affiliation{Department of Physics, National Tsing Hua University, 101, Section 2. Kuang-Fu Road, Hsinchu, 30013, Taiwan (R.O.C.)}

\begin{abstract}
We report the properties of molecular gas in a sample of six host galaxies of fast radio bursts (FRBs)
obtained from CO observations with the Atacama Large Millimeter/submillimeter Array (FRBs 20180924B, 20190102C, and 20190711A) and results of one non-detection in a dwarf galaxy (FRB\,20121102A) and two events detected in M81 (FRB\,20200120E) and the Milky Way (FRB\,20200428A). 
The CO observations resulted in the detection of CO(3--2) emission in the FRB\,20180924B host and non-detections of CO(3--2) and CO(2--1) emission in the hosts of FRB\,20190102C and FRB\,20190711A, respectively. 
The derived molecular gas mass and 3$\sigma$ upper limit is 
$(2.4 \pm 0.2) \times 10^9$ $M_{\odot}$,
$<$$3.8 \times 10^8$ $M_{\odot}$, and 
$<$$6.7 \times 10^9$ $M_{\odot}$
for the hosts of FRB\,20180924B, FRB\,20190102C, and FRB\,20190711A, respectively. 
We found diversity in molecular gas properties (gas mass, gas depletion time, and gas fraction to stellar mass) in the sample. 
Compared to other star-forming galaxies, 
the FRB\,20180924B host is gas-rich (the larger molecular gas fraction), and 
the hosts of FRB\,20190102C and FRB\,20200120E are gas-poor with a shorter depletion time for their stellar mass and star-formation rate. 
Our findings suggest that FRBs arise from multiple progenitors or single progenitors that can exist in a wide range of galaxy environments. 
Statistical analysis shows a significant difference in the distribution of molecular gas fraction between the FRB hosts and local star-forming galaxies. However, the difference is not substantial when an outlier, the FRB\,20200120E host, is excluded, and analysis with a larger sample is needed. 

\end{abstract}

\keywords{
\href{http://astrothesaurus.org/uat/2008}{Radio transient sources (2008)}; 
\href{http://astrothesaurus.org/uat/1339}{Radio bursts (1339)}; 
\href{http://astrothesaurus.org/uat/508}{Extragalactic radio sources (508)}; 
\href{http://astrothesaurus.org/uat/573}{Galaxies (573)};
\href{http://astrothesaurus.org/uat/262}{CO line emission (262)};
\href{http://astrothesaurus.org/uat/1073}{Molecular gas (1073)};
\href{http://astrothesaurus.org/uat/847}{Interstellar medium (847)};
\href{http://astrothesaurus.org/uat/1569}{Star formation (1569)};
\href{http://astrothesaurus.org/uat/1338}{Radio astronomy (1338)}; 
\href{http://astrothesaurus.org/uat/1346}{Radio interferometry (1346)}; 
\href{http://astrothesaurus.org/uat/1145}{Observational astronomy (1145)};
}

\section{Introduction}

Fast radio bursts (FRBs) are bright, energetic radio pulses with a duration of microsecond--millisecond \citep[][for a review]{cord19}. 
FRBs' physical mechanism and progenitors are one of the biggest mysteries of modern astronomy. 
FRBs are now thought to be extragalactic phenomena because of their large dispersion measures \citep[e.g.,][]{cord19} together with the identifications of extragalactic host galaxies \citep[e.g.,][]{hein20a}. 
FRBs are classified into repeating bursts and (apparently) non-repeating (one-off) bursts. 
It is highly debated whether the two types of FRBs are the same population or not \citep[e.g.,][]{hash20, pleu21}. 
Many theoretical models have been proposed to explain FRBs \citep{plat19}. 
Recently, a millisecond-duration radio burst (FRB\,20200428A) was detected from the Galactic magnetar SGR\,1935+2154 \citep{the20, boch20a}, suggesting a magnetar origin of other extragalactic FRBs.

Because the link between progenitors and specific galactic environments is expected, observations of host galaxies are essential to understand their nature and constrain models, as demonstrated by studies of astronomical transients such as gamma-ray bursts \citep[GRBs; e.g.,][]{lef06, perl16a, niin17, hash19b}. %
Only recently, new facilities and extensive follow-up campaigns enable localization of FRBs and identifications of host galaxies \citep[e.g.,][]{chat17, bann19, proc19b, macq20, marc20}. 
Optical/near-infrared observations show that the hosts span a wide range in morphology, color, stellar mass, and star-formation rate (SFR) \citep[e.g.,][]{hein20a, bhan22}. 
Comparison with the hosts of other transients found that the generally massive and metal-rich environments of FRBs disfavor similar progenitor formation channels to those of long-duration GRBs (LGRBs) and superluminous supernovae (SLSNe) \citep[e.g.,][]{li19, li20b, bhan20, hein20a}.

CO observations of host galaxies provide physical properties of the interstellar medium as a material for star formation in the environment of FRBs, such as molecular gas content, gas depletion time, gas fraction, and star-formation efficiency. 
\cite{bowe18} conducted Submillimeter Array and Northern Extended Millimeter Array observations of the host galaxy of FRB\,20121102A in the CO(3--2) and CO(1--0) transitions, respectively, and did not detect emission from either transition.

In this Letter, we present molecular gas observations in the host galaxies of three FRBs (20180924B, 20190102C, and 20190711A) performed with the Atacama Large Millimeter/submillimeter Array (ALMA). 
Adding the other three host galaxies of FRBs (20121102A, 20200120E, and 20200428A) for which molecular gas data are available in the literature, we discuss the properties of molecular gas in the six host galaxies.
Section~\ref{sec:alma} describes ALMA observations of the FRB host galaxies, data reduction, and results.
In Section~\ref{sec:discussion}, we derive the physical properties of molecular gas in the FRB hosts, compare them with those of other galaxy populations, and discuss environments where FRB progenitors originate.
Conclusions and implications for progenitors are presented in Section~\ref{sec:conclusions}. 
Throughout the paper, we adopt a \cite{chab03} initial mass function and cosmological parameters of $H_0=67.7$ km s$^{-1}$ Mpc$^{-1}$ and $\Omega_{\rm{M}}=0.310$ based on the {\sl Planck} 2018 results \citep{plan20f}. 


\section{ALMA Data} \label{sec:alma}

\subsection{Target Host Galaxies}

FRB\,20180924B was detected on 2018 September 24 at 16:23:12.6265 UTC with the Australian Square Kilometre Array Pathfinder \citep[ASKAP;][]{john08} and was localised to a host galaxy at $z = 0.3214$, DES J214425.25$-$405400.81 \citep{bann19}. 
The host has the stellar mass of $M_* = (13.2 \pm 5.1) \times 10^9$ $M_\odot$, the SFR of $0.88 \pm 0.26$ $M_\odot$~yr$^{-1}$, and a gas-phase metallicity of $12 + \log(\rm O/H) = 8.93^{+0.02}_{-0.02}$  \citep{hein20a}. 
It is located between star-forming and quiescent galaxies (``green valley'') in a color-magnitude diagram and an SFR--$M_*$ plot \citep{hein20a}. 
A Baldwin, Phillips \& Terlevich (BPT) diagram \citep{bald81} shows that the host has a characteristic of low-ionization narrow-emission line region (LINER) \citep{bann19}.

FRB\,20190102C was detected on 2019 January 2 at 05:38:43.49184 UTC with ASKAP and was localised to a host galaxy at $z = 0.291$ \citep{macq20}. 
The host has the stellar mass of $(3.39 \pm 1.02) \times 10^9$ $M_\odot$, the SFR of $0.86 \pm 0.26$ $M_\odot$~yr$^{-1}$, and the metallicity of $12 + \log(\rm O/H) = 8.70^{+0.08}_{-0.07}$ \citep{hein20a}. 
The host is located in the ``composite'' region close to the boundary of star-forming galaxies and active galactic nuclei (AGNs) on the BPT diagram \citep{bhan20}.

FRB\,20190711A was detected on 2019 July 11 at 01:53:41.09338 UTC with ASKAP and was localized to a host galaxy at $z = 0.5220$ \citep{macq20}. 
The host galaxy is a typical star-forming galaxy at its redshift with the stellar mass of $(0.81 \pm 0.29) \times 10^9$ $M_\odot$ and the SFR of $0.42 \pm 0.12$ $M_\odot$~yr$^{-1}$ \citep{hein20a}.

\subsection{Observations and Data Reduction}

We conducted ALMA Band 4 CO(2--1) observations of the FRB\,20190711A host from May 2 to June 09, 2022 (Project code: 2021.1.00027.S). 
The correlator was used in the time domain mode with a bandwidth of 1.875 GHz (488.28 kHz $\times$ 3840 channels). 
Four basebands were used, providing a total bandwidth of 7.5 GHz. 
The array configuration was C-3 and C-4 with the baseline lengths of 15.1--500.2 and 15.1--783.5 m, respectively. 
The number of the available antenna was 41--46, and the on-source integration time was 6.95 hours. 
Bandpass and flux calibrations were performed with J2357$-$5311 and phase calibrations with J2116$-$8053.

We utilized the ALMA archive data of Band 6 CO(3--2) observations of the hosts of FRBs 20180924B and 20190102C (Project code: 2019.01450.S). 
Observations of the FRB\,20180924B host were conducted on November 26, 2019, using 44--46 antennas in an array configuration with baseline lengths of 15.0--313.7 m. 
The correlator setup consists of a baseband with the bandwidth of 1.875 GHz (7812.5 MHz $\times$ 240 channels) for the CO line and three basebands with the bandwidth of 2 GHz (15.625 MHz $\times$ 128 channels) for the continuum. 
The on-source integration time was 27 minutes.
Observations of the FRB\,20190102C host were conducted on November 30 and December 03, 2019, using 40--43 antennas in an array configuration with baseline lengths of 15.1--312.7 m. 
The same correlator setup was used for the FRB\,20180924B host observations. 
The on-source integration time was 88 minutes.

The data were reduced with Common Astronomy Software Applications \citep[CASA;][]{mcmu07}. 
Data calibration was done with the ALMA Science Pipeline Software of CASA. 
Maps were processed with a \verb|tclean| task with Briggs weighting and a \verb|robust| parameter of 2.0. 
A clean box was placed for imaging the CO emission of the FRB\,20180924B host, where a component with a peak signal-to-noise ratio (S/N) above five was identified. 
The continuum maps were created by excluding channels with CO emission. 
The RMS noise level of the CO maps is 34, 340, and 67 $\mu$Jy~beam$^{-1}$ with a velocity resolution of 50 km~s$^{-1}$  for the hosts of FRBs 20180924B, 20190102C, and 20190711A, respectively. 
The RMS noise level of the continuum maps is 47, 56, and 3.8 $\mu$Jy~beam$^{-1}$.

\begin{figure*}
\gridline{
\fig{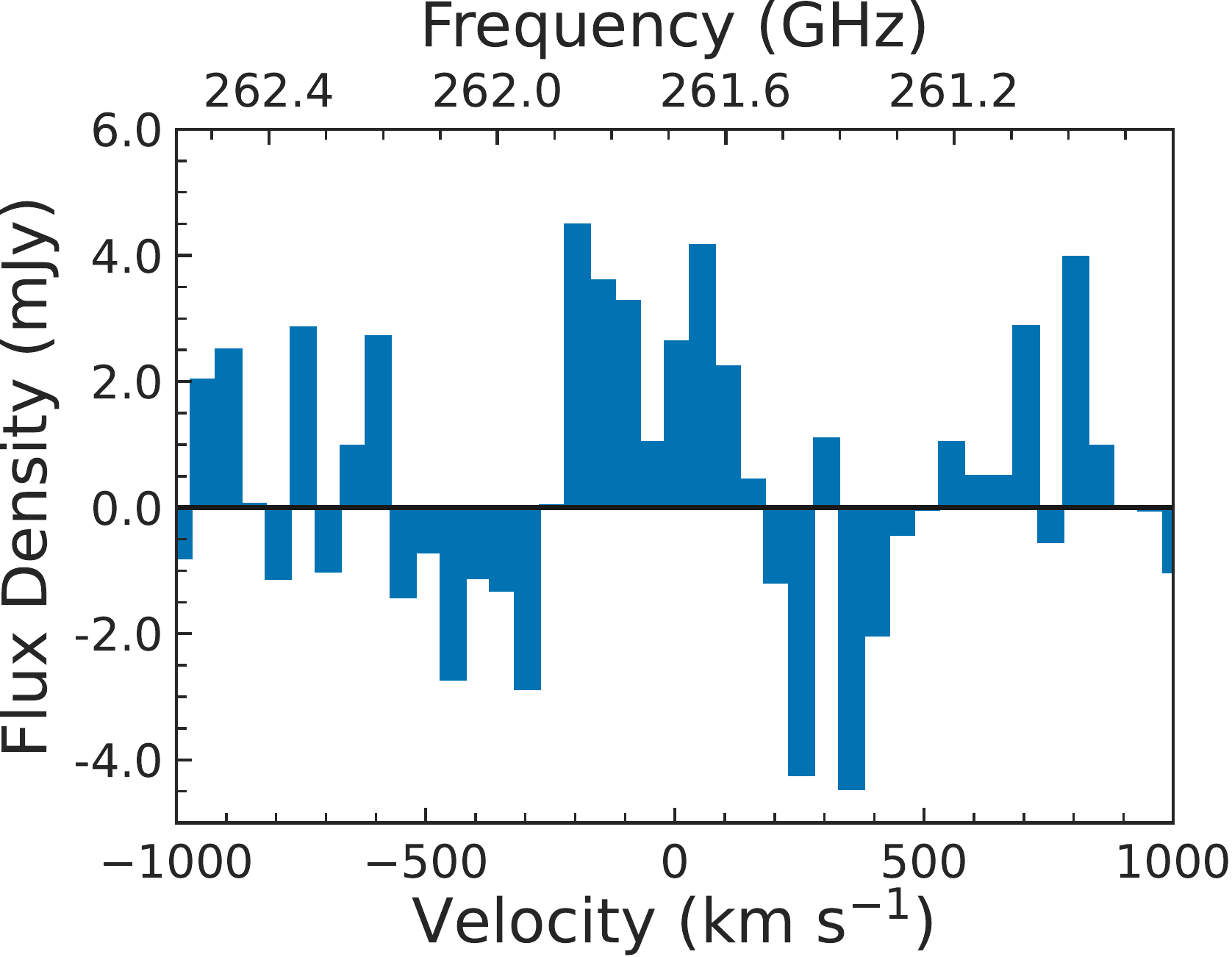}{0.22\textheight}{}
\fig{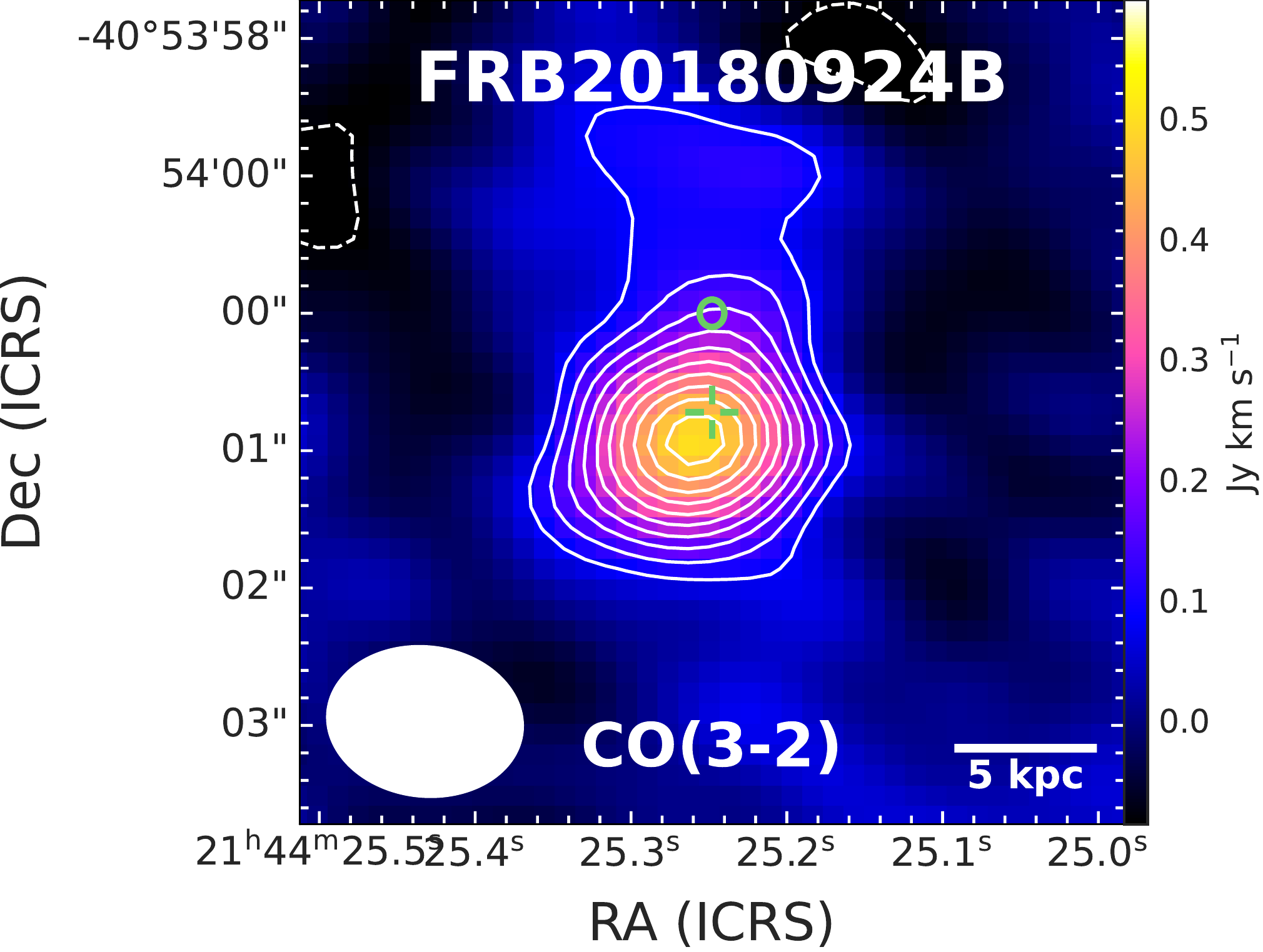}{0.3\textwidth}{}
\fig{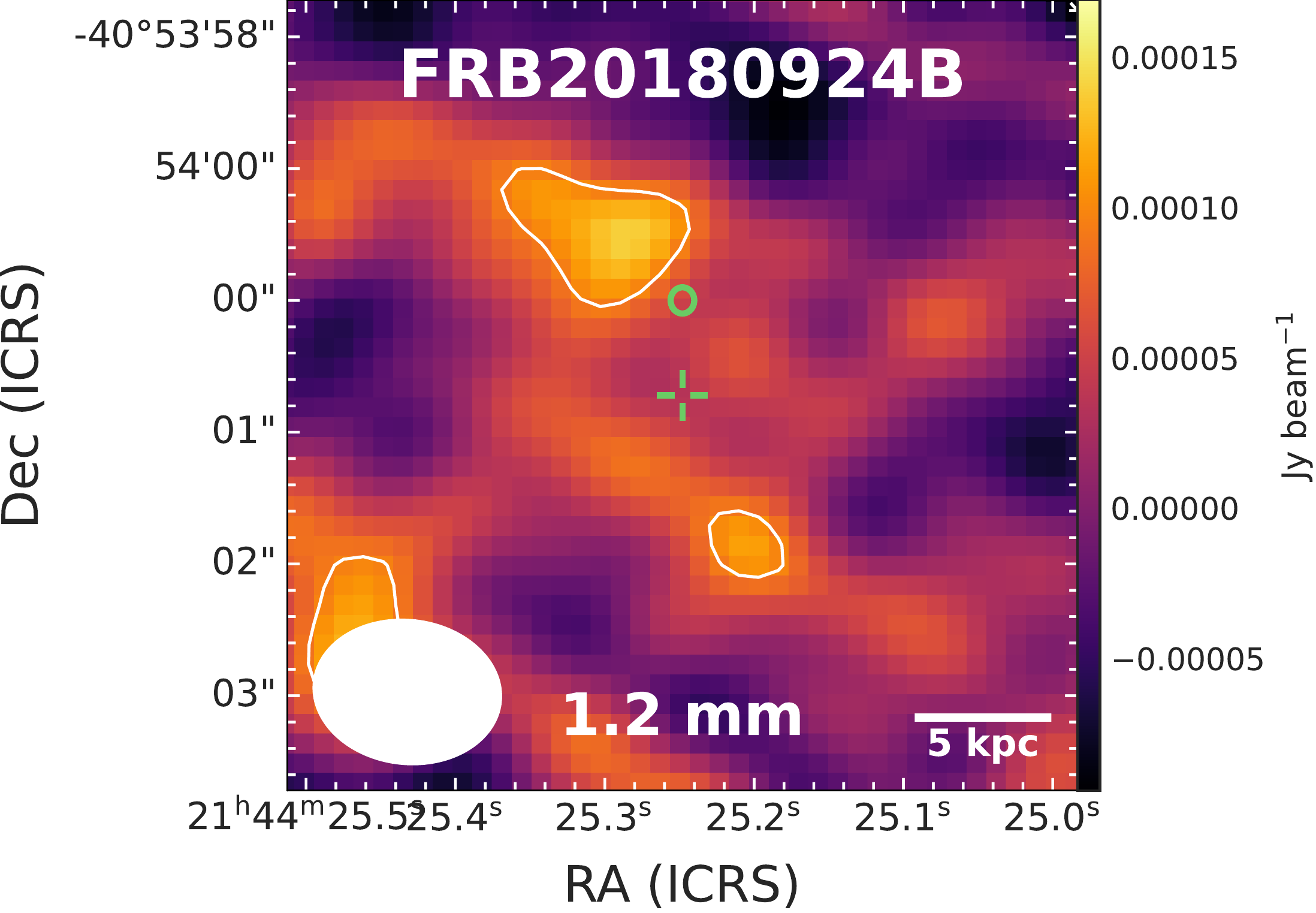}{0.32\textwidth}{}
}
\gridline{
\fig{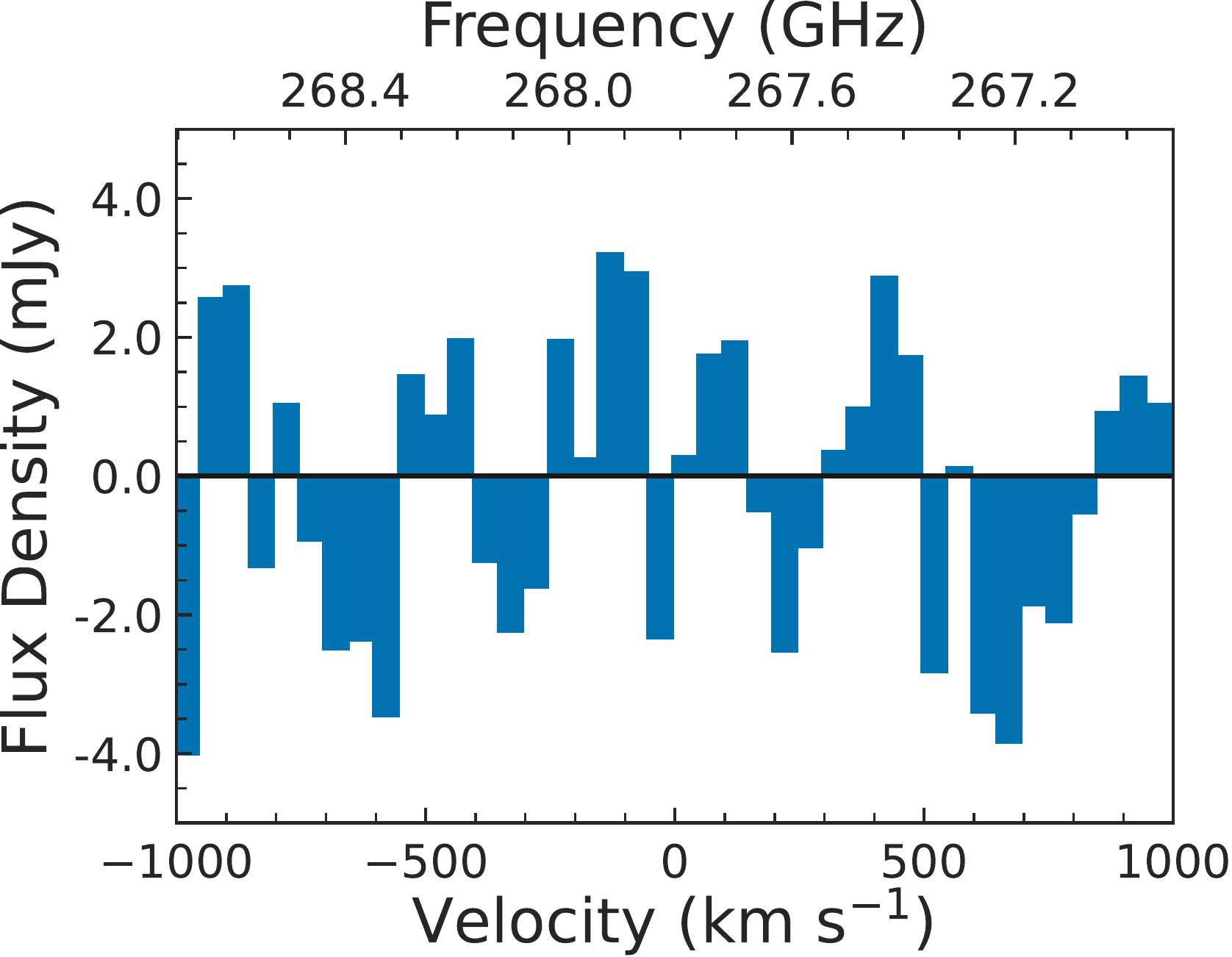}{0.22\textheight}{}
\fig{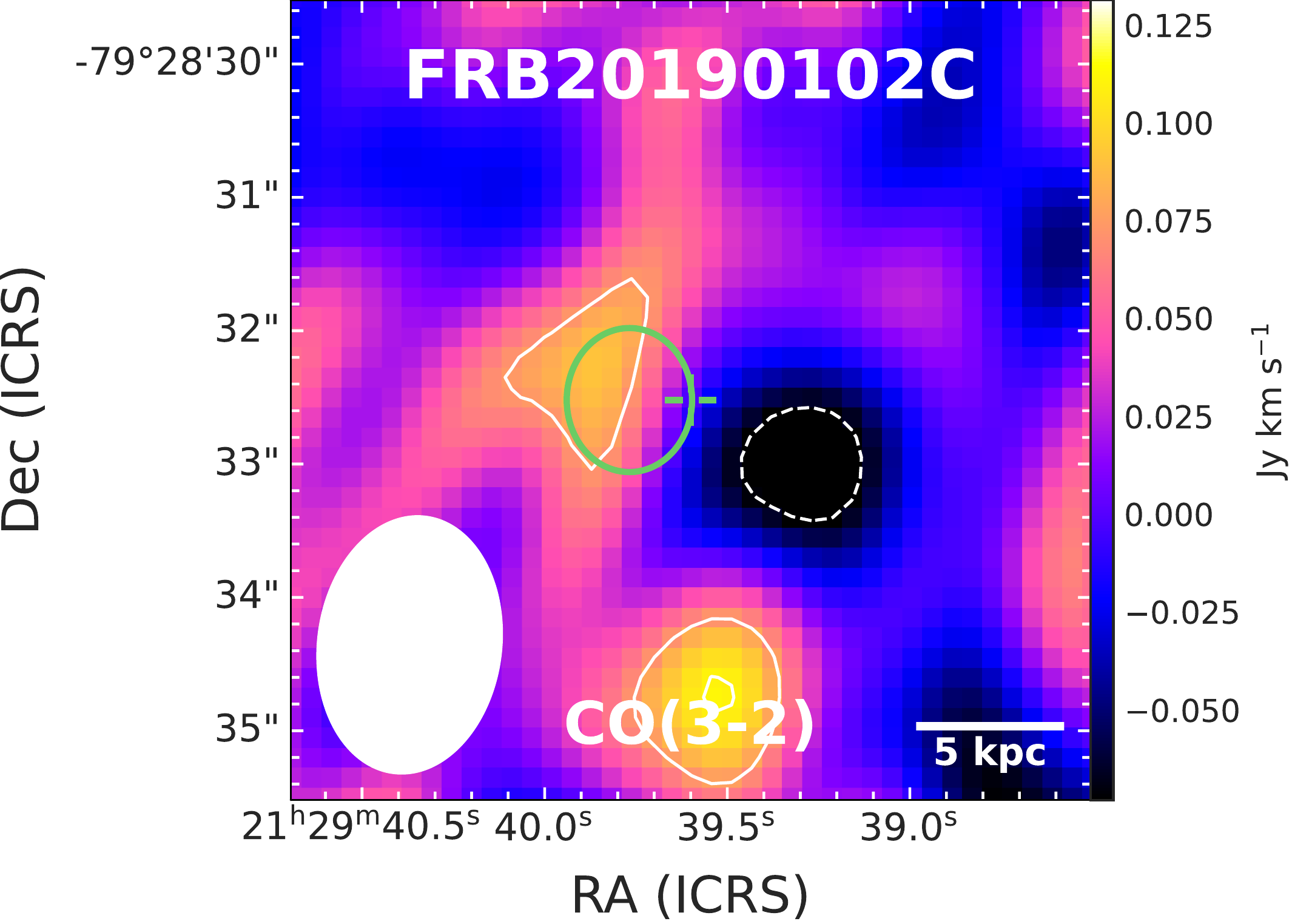}{0.32\textwidth}{}
\fig{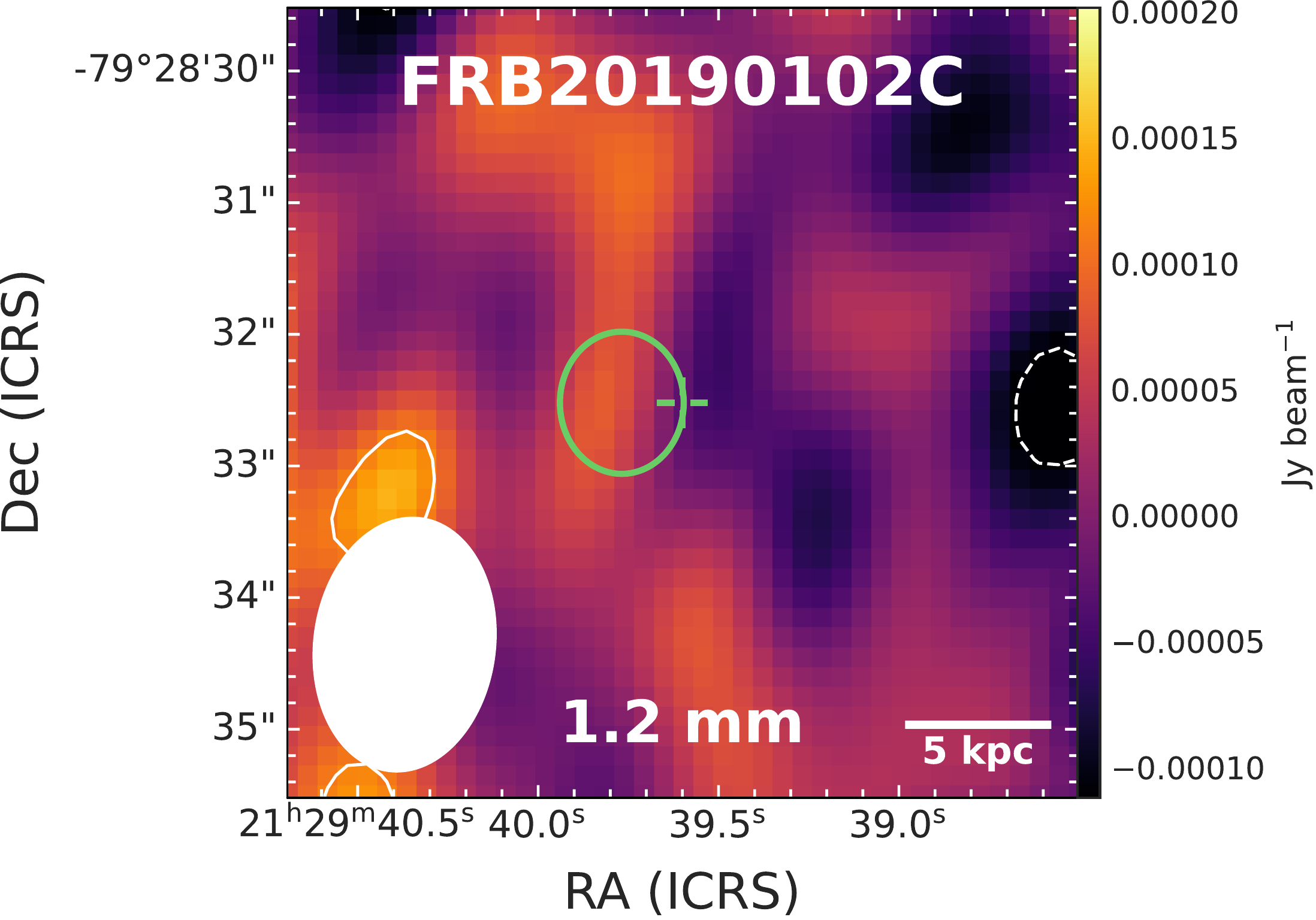}{0.32\textwidth}{}
}
\gridline{
\fig{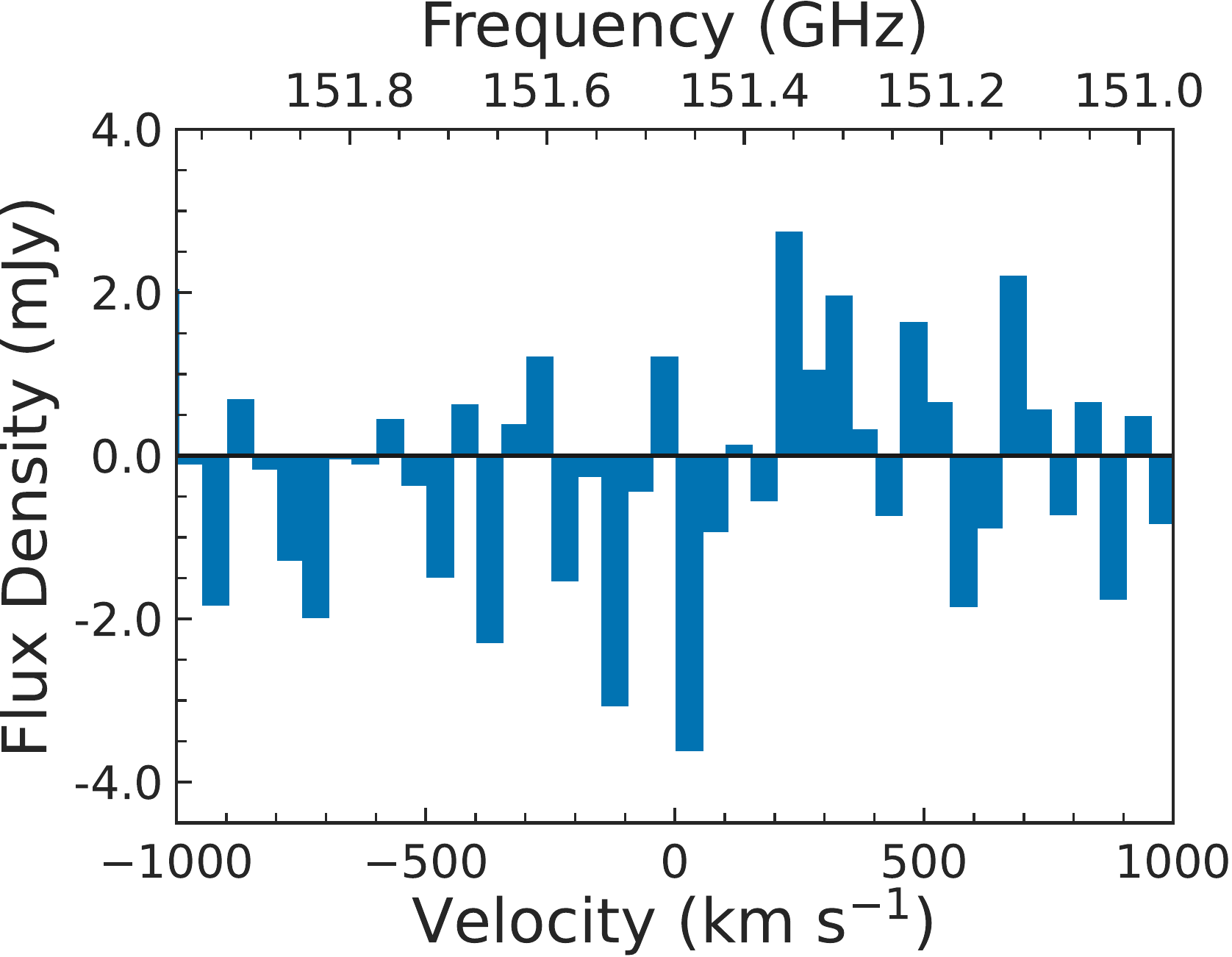}{0.22\textheight}{}
\fig{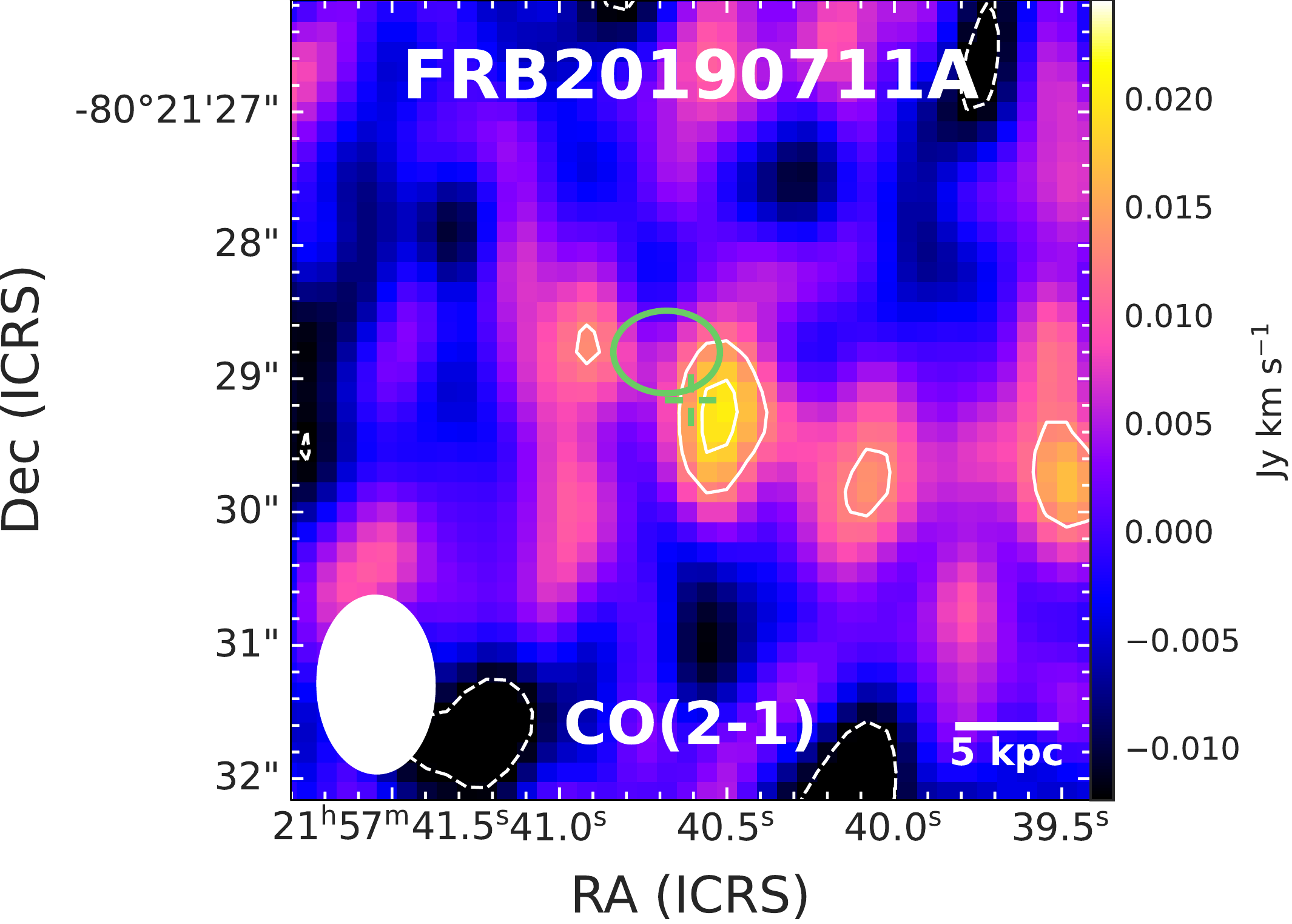}{0.32\textwidth}{}
\fig{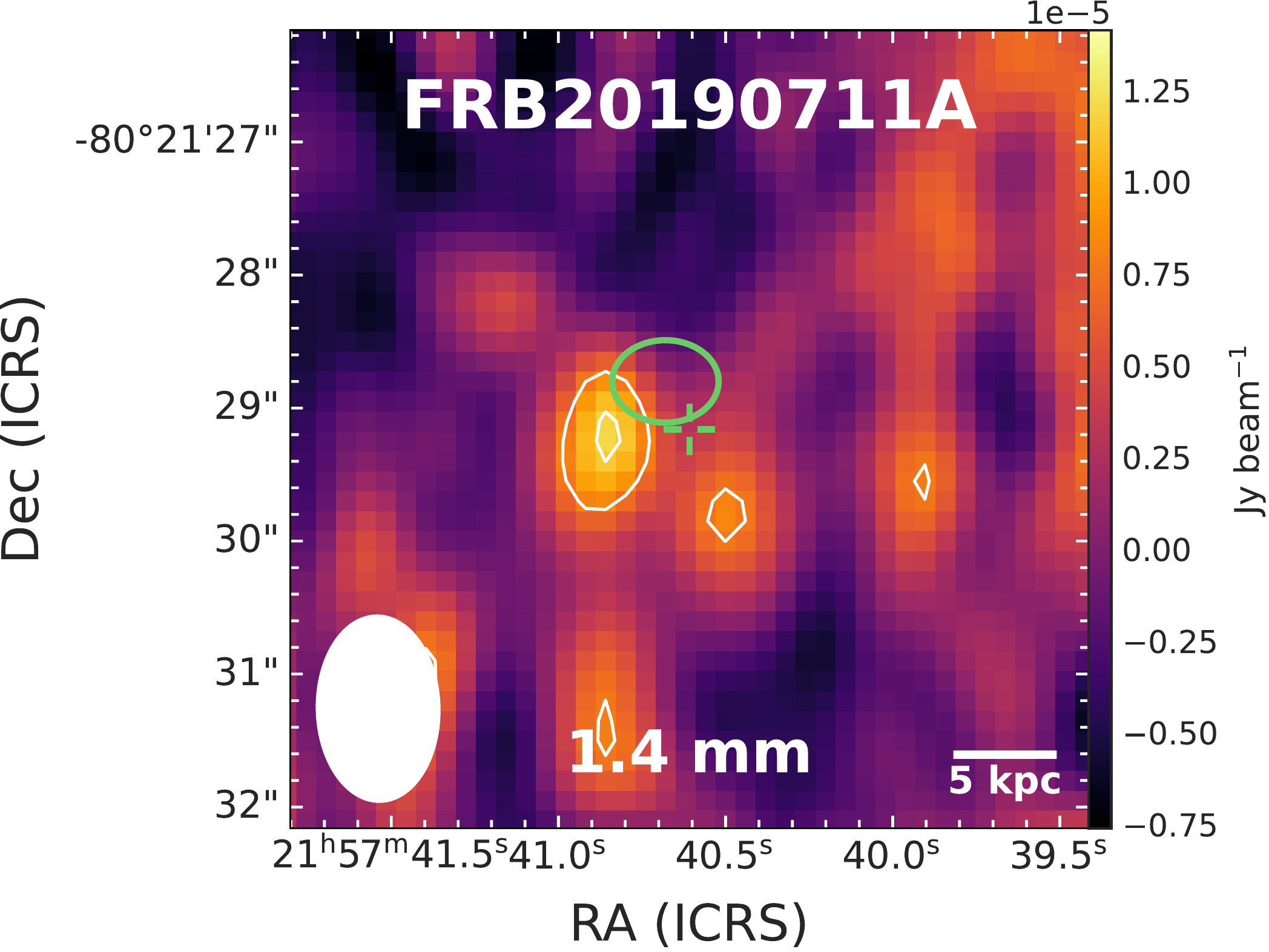}{0.32\textwidth}{}
}
\caption{
CO spectrum (left column), CO velocity-integrated intensity map (middle column), and continuum map (right column) of the host galaxies of FRB\,20180924B (top row), FRB\,20190102C (middle row), and FRB\,20190711A (bottom row). 
The velocity resolution of the spectrum is 50 km~s$^{-1}$, and the line flux is measured with a $1''$-radius aperture centered at the host galaxy position. 
The velocity from $-200$ to $+$200 km~s$^{-1}$ is integrated for the velocity-integrated intensity maps of the hosts of FRBs 20180924B and 20190102C, while the range from $+200$ to $+$400 km~s$^{-1}$ is integrated for the FRB\,20190711A host. 
The contours are $\pm$$2\sigma$, $\pm$$3\sigma$, and a $1\sigma$ step subsequently. 
The maximum contour level for the velocity-integrated intensity map of the FRB20180924A host is 11$\sigma$.
The green circle shows the 90\% confidence region of the FRB position \citep{macq20}.
The green cross represents the host position \citep{hein20a}. 
The synthesized beam size is shown in the lower-left corner of the maps. 
}
\label{fig:results}
\end{figure*}

\begin{table*}[t]
\caption{Results of CO ALMA Observations} \label{tab:results}
\begin{tabular}{ccccCC}
\hline\hline
FRB & CO$^{\rm a}$ & Beam size & RMS$^{\rm b}$ & 
$S_{\rm CO}\Delta v^{\rm c}$ & ${L'}_{\rm CO}{}^{\rm d}$ \\
 & & (arcsec) & \colhead{($\mu$Jy beam$^{-1}$)} & 
\colhead{(Jy km s$^{-1}$)} & \colhead{(K km s$^{-1}$ pc$^2$)} \\
\hline
20180924B & 3--2 & $1\farcs43 \times 1\farcs09$ (PA $= +83.7^{\circ}$) & 34 & 0.50\pm0.04 & (3.1\pm0.3)\times10^8 \\
20190102C & 3--2 & $1\farcs94 \times 1\farcs37$ (PA $= -7.2^{\circ}$) & 340 &
<0.088 & <4.9\times10^7 \\ 
20190711A & 2--1 & $1\farcs33 \times 0\farcs88$ (PA $= +0.8^{\circ}$) & 67 & 
<0.017 & <7.3\times10^7 \\ 
\hline
\end{tabular}
\tablecomments{
Limits are 3$\sigma$. \\
$^{\rm a}$ CO rotational transition. \\
$^{\rm b}$ RMS noise level with a velocity resolution of 50 km s$^{-1}$. \\
$^{\rm c}$ Velocity-integrated CO flux. A velocity width of 160 km s$^{-1}$ is assumed for nondetection. \\
$^{\rm d}$ Line luminosity of CO(3--2) or CO(2--1). \\
}
\end{table*}

\subsection{Results} \label{sec:results}

Figure~\ref{fig:results} shows CO spectrum, velocity-integrated CO maps, and continuum maps. 
The CO(3--2) emission is significantly detected in the FRB\,20180924B host with a peak S/N of 12 in the velocity-integrated intensity map. 
The emission is spatially unresolved with the synthesized beam. 
The peak position of the CO emission coincides with the host galaxy center. 
The CO spectra shows a full width at zero intensity of $\sim$400 km~s$^{-1}$ and a full width at half maximum of $\sim$270 km~s$^{-1}$, which is comparable to those of local star-forming galaxies with $M_* > 10^{10}$ $M_\odot$ \citep{sain17}. 
In the FRB\,20190711A host spectrum, there is a feature of emission at 200--400 km~s$^{-1}$. 
Although the peak position of the velocity-integrated intensity map coincides with the host galaxy center, the feature is marginal (S/N $= 3.3$). 
There is also a 3.1$\sigma$ signal in the continuum map of the FRB\,20190711A host, although the peak position is $\sim$0\farcs6 away from the host galaxy center. 
Except for the CO emission in the FRB\,20180924B host, we do not find significant emission in the CO line and continuum maps.

\section{Discussion} \label{sec:discussion}

\subsection{Molecular Gas Mass} \label{sec:gas}

We derive molecular gas mass from the results of CO observations. 
The CO luminosity is calculated as follows \citep{solo05}, 
\begin{eqnarray}
L'_{\rm CO} = 3.25 \times 10^7 S_{\rm CO}\Delta v \nu_{\rm obs}^{-2} D_{\rm L}^2 (1+z)^{-3}, 
\end{eqnarray}
where $L'_{\rm CO}$ is in K km s$^{-1}$~pc$^2$, $S_{\rm CO}\Delta v$ is the velocity-integrated intensity in Jy~km~s$^{-1}$, $\nu_{\rm obs}$ is the observed line frequency in GHz, and $D_{\rm L}$ is the luminosity distance in Mpc. 
The molecular gas mass is derived from 
\begin{eqnarray}
M_{\rm gas} = \alpha_{\rm CO} L'_{\rm CO(1-0)}. 
\end{eqnarray}
The CO(1--0) luminosity, $L'_{\rm CO(1-0)}$, is derived by assuming CO line luminosity ratios of CO(2--1)/CO(1--0) = 0.65 and CO(3--2)/CO(1--0) = 0.55, which are typical values for local star-forming galaxies \citep[e.g.,][]{sain17, lamp20, yaji21, den21}. 
The conversion factor $\alpha_{\rm CO}$ is thought to be dependent on gas-phase metallicity, increasing $\alpha_{\rm CO}$ with decreasing metallicity \citep[e.g.,][]{bola13}.
The Milky Way-like $\alpha_{\rm CO}$ of 4.3 $M_{\odot}$~(K~km~s$^{-1}$~pc$^2$)$^{-1}$ \citep{bola13} is adopted for the hosts of FRBs 20180924B and 20190102C, where the gas-phase metallicity is comparable to that of the Milky Way. 
Because the metallicity of the FRB\,20190711A host was not obtained from emission line diagnostics, we derive the metallicity of $12+\log(\rm O/H) = 8.3$ from the mass--metallicity relation of \cite{genz15}. 
The metallicity is lower than that of the Milky Way, and we adopt a metallicity-dependent $\alpha_{\rm CO}$. 
There are many previous studies of the relation between metallicity and $\alpha_{\rm CO}$ \citep[e.g.,][]{wils95, genz12, schr12, bola13}. 
The relation of \cite{genz12}, where they use the samples of local and high-redshift star-forming galaxies, gives $\alpha_{\rm CO}$ of $8.9 M_{\odot}$~(K~km~s$^{-1}$~pc$^2$)$^{-1}$. 
If we adopt the relation of \cite{schr12} for non-starburst galaxies, where they use the samples of nearby star-forming galaxies including low-metallicity dwarf galaxies, gives $\alpha_{\rm CO}$ of $60~M_{\odot}$~(K~km~s$^{-1}$~pc$^2$)$^{-1}$. 
To conservatively estimate an upper limit on molecular gas mass, we adopt $\alpha_{\rm CO} = 60~ M_{\odot}$~(K~km~s$^{-1}$~pc$^2$)$^{-1}$ for the FRB\,20190711A host. 
The molecular gas mass of the FRB\,20180924B host is estimated to be $(2.4 \pm 0.2) \times 10^9$ $M_{\odot}$. 
The 3$\sigma$ upper limit for the hosts of FRBs 20190102C and 20190711A is $<${}$3.8 \times 10^8$ $M_{\odot}$, and $<${}$1.2 \times 10^{10}$ $M_{\odot}$ respectively, by assuming a line width of 160 K~km~s$^{-1}$, which is an average CO line width of nearby low-mass ($10^{8.5} < M_*/M_\odot < 10^{10}$) star-forming galaxies in the ALLSMOG sample \citep{cico17}. 
The results of ALMA observations on molecular gas are summarized in Table~\ref{tab:results}.

\subsection{Molecular Gas Properties} \label{sec:properties}

In addition to the FRB hosts observed by ALMA, we include three FRB hosts for which molecular gas information is available in the literature in the discussion to expand the sample size.
The CO(1--0) and (3--2) observations of the FRB\,20121102A host by \cite{bowe18} provided 3$\sigma$ upper limits of $L'_{\rm CO(1-0)} < 2.3 \times 10^9$ K km s$^{-1}$~pc$^2$ and $L'_{\rm CO(3-2)} < 2.5 \times 10^8$ K km s$^{-1}$~pc$^2$. 
The molecular gas mass of $M_{\rm gas} < 9.1 \times 10^9$ $M_\odot$ is derived from the CO(3--2) line luminosity by using the metallicity-dependent $\alpha_{\rm CO}$ of $200~M_{\odot}$~(K~km~s$^{-1}$~pc$^2$)$^{-1}$ \citep{schr12} and the same CO line luminosity ratio adopted for our sample. 
We also include two events: FRB\,20200120E and FRB\,20200428A. 
FRB\,20200120E is the closest known extragalactic FRB, localized to a globular cluster associated with M81 \citep{bhar21, kirs22}. 
FRB\,20200428A was identified as the Galactic magnetar SGR\,$1935+2154$ \citep{the20, boch20a}. 
Since they occupy different positions on the phase space of luminosity and duration of FRBs, it is actively debated whether they are the same populations as other FRBs \citep[e.g.,][]{nimm22}. 
Although the positions of the two nearby FRBs within the host galaxies are localized, we use the values for the entire host galaxies to treat them the same way as other FRBs. 
The host galaxy properties are presented in Table~\ref{tab:properties}.

\begin{figure}
\centering
\includegraphics[width=\linewidth]{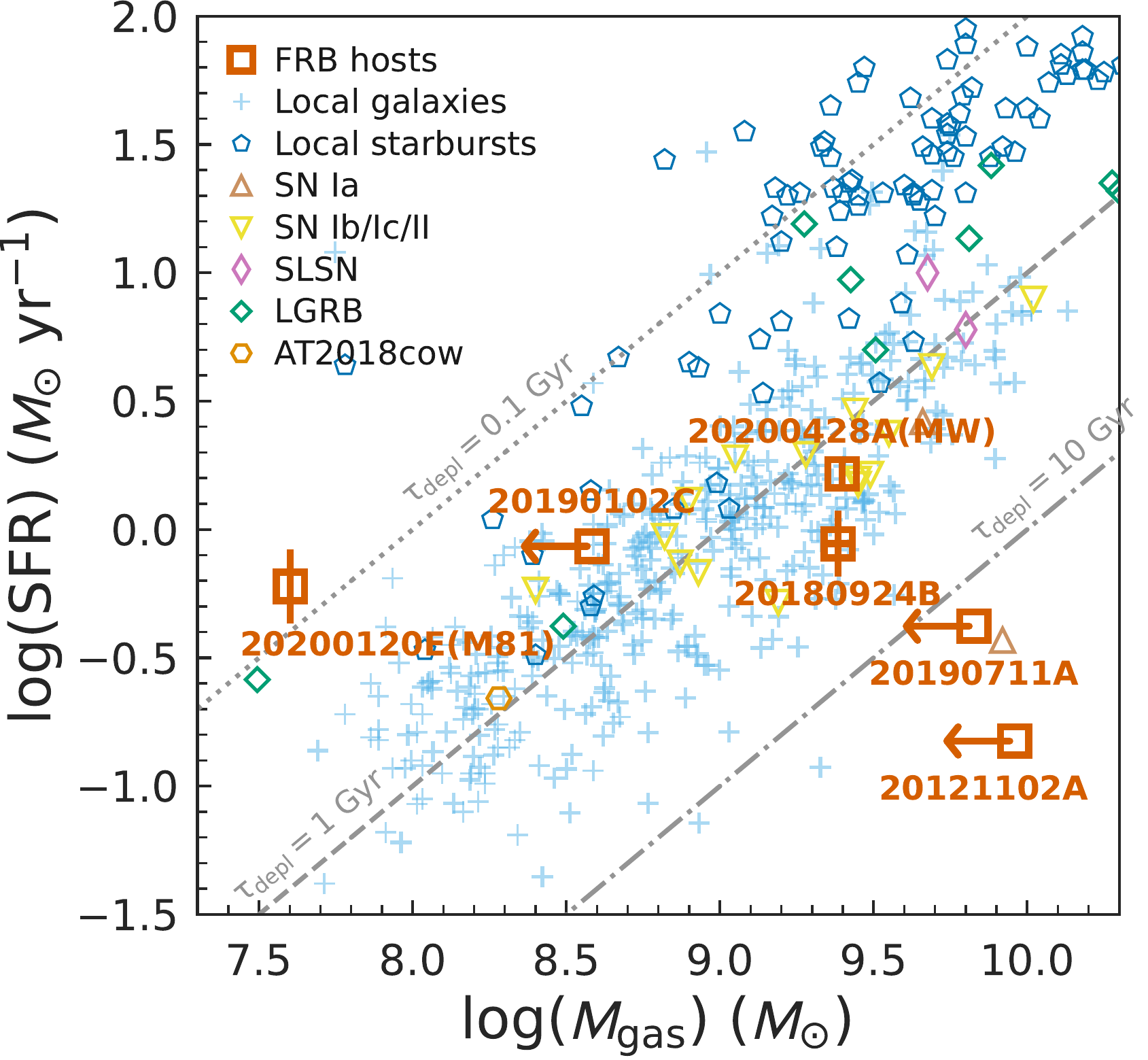}
\caption{
Comparison of molecular gas mass and SFR of the FRB hosts. 
For comparison, we plot local star-forming galaxies from the samples of ALLSMOG \citep{cico17} and xCOLDGASS \citep{sain17}, starbursts \citep{kenn21}, 
the hosts of SNe \citep{galb17}, SLSNe \citep{hats20, arab19}, LGRBs \citep{hats20a}, and AT2018cow \citep{moro19}. 
The dot, dash, and dot-dash lines represent gas depletion times of 0.1, 1, and 10 Gyr, respectively. 
}
\label{fig:mgas-sfr}
\end{figure}

We compare the molecular gas mass of the FRB hosts with SFR in Figure~\ref{fig:mgas-sfr}. 
We find that the FRB hosts have diversity in molecular gas properties. 
The hosts span a wide range in the plot with the molecular gas mass from $4 \times 10^7$ $M_\odot$ to $2.4\times10^9$ $M_\odot$ and the gas depletion time from 0.07 to 2.8 Gyr. 
While the FRB\,20180924B host is molecular gas-rich compared to local star-forming galaxies, the hosts of FRBs 20190102C and 20200120E are molecular gas-poor for its SFR with a shorter gas depletion time.

\begin{figure*}
\centering
\plottwo{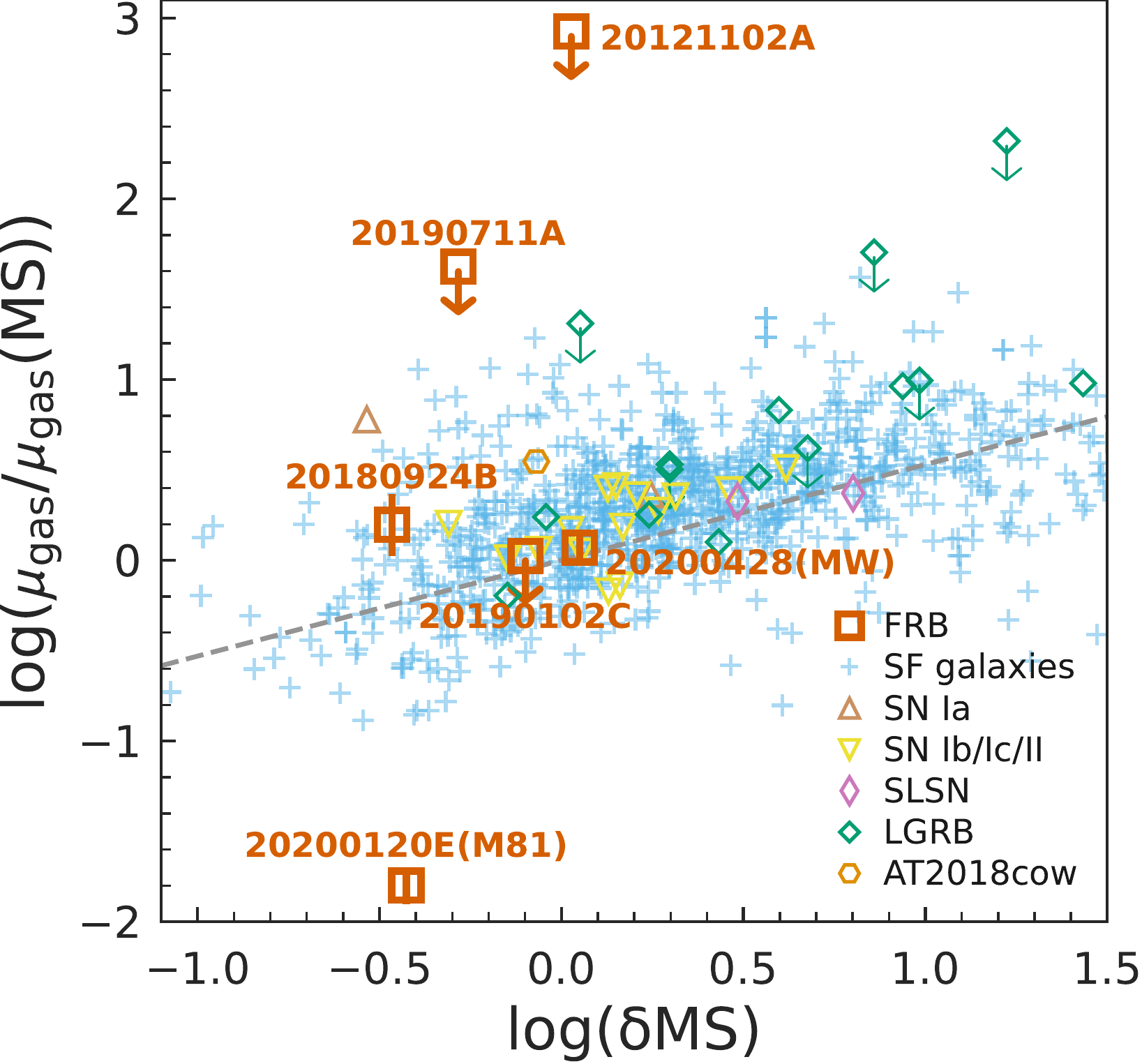}{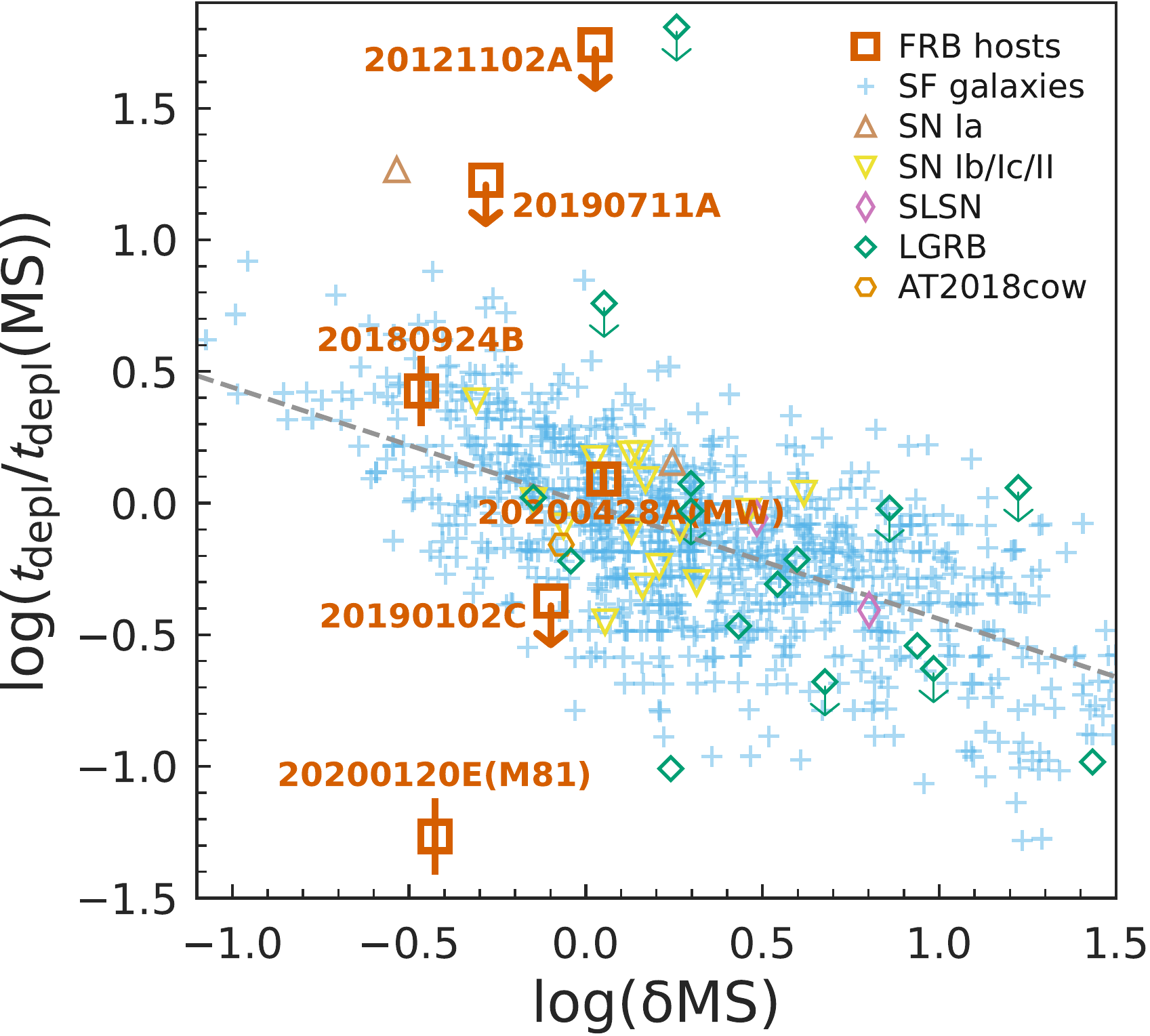}
\caption{
Molecular gas fraction (left) and gas depletion time (right) as a function of the offset from the reference MS line after removal of redshift dependence. 
For comparison, star-forming galaxies \citep{cico17, tacc18} and the hosts of SNe \citep{galb17}, SLSNe \citep{hats20, arab19}, LGRBs \citep{hats20a}, and AT2018cow \citep{moro19} are presented. We restricted the redshift of the galaxies in the comparison to $z < 1$ to match the redshift range of the FRB hosts.
The dashed line shows the best-fit line for the MS galaxies derived in \cite{tacc18}. 
}
\label{fig:dms}
\end{figure*}

The molecular gas mass fraction and gas depletion time depend on redshift, stellar mass, and offset from the main-sequence (MS) line \citep[e.g.,][]{genz15}. 
\cite{genz15} and \cite{tacc18} argue that the parameter dependencies of $\mu_{\rm gas}$ and $\tau_{\rm gas}$ can be separated as products of power laws of redshift, offset from the MS line ($\delta$MS $=$ sSFR/sSFR(MS, $z$, $M_*$)), and stellar mass: 
$\mu_{\rm gas} \propto (1 + z)^{2.5} (\delta{\rm MS})^{0.52} (M_*)^{-0.36}$, 
$\tau_{\rm gas} \propto (1 + z)^{-0.6} (\delta{\rm MS})^{-0.44}$. 
In Figure~\ref{fig:dms}, we show $\mu_{\rm gas}$ and $\tau_{\rm gas}$ as a function of $\delta$MS after removing the above redshift dependence for each galaxy.
We find that the FRB\,20180924B host has a larger molecular gas fraction, and the FRB\,20190102C host has a shorter gas depletion time for their stellar mass and SFR, deviating with $>$0.4 dex from the scaling relations of other star-forming galaxies.
The FRB\,20200120E host has a smaller gas fraction and a shorter gas depletion time and is largely offset from the distributions of other star-forming galaxies. 
Those FRB hosts are offset from the scaling relations of other star-forming galaxies, which is in contrast to the majority of the hosts of core-collapse SNe, SLSNe \citep{hats20}, and LGRBs \citep{hats20a}, suggesting different origins of FRBs. 
The small sample size also prevents us from discussing the difference between repeating and non-repeating FRBs, leaving it for future works.

To discuss the statistical significance of the difference between the FRB hosts and normal star-forming galaxies, we compare with the ALLSMOG sample \citep{cico17}. 
The sample consists of star-forming galaxies at $0.01 < z < 0.03$ with stellar masses in the range $8.5 < \log{(M_*/M_\odot)} < 10$, which matches to the mass range of our FRB host sample. 
Because the redshift distribution differs between the two samples, we remove the redshift dependence of $\mu_{\rm gas}$ and $\tau_{\rm depl}$ by using the prescription of \cite{tacc18} as stated above. 
To take into account non-detections in the samples, we adopt the Kaplan-Meier estimator \citep{kapl58}, a non-parametric statistic to estimate the probability distribution of data containing non-detections. 
We applied this analysis to $\mu_{\rm gas}$ and $\tau_{\rm depl}$
using a Python package \verb|lifelines| \citep{davi19}.
The log-rank test shows a $p$-value of 0.048 and 0.83 for $\mu_{\rm gas}$ and $\tau_{\rm depl}$, respectively, suggesting a difference in the distribution of $\mu_{\rm gas}$ between the two samples. 
If we remove an outlier, the FRB\,20200120E host, it becomes inconclusive with a $p$-value of 0.19, and a larger sample is required to test the difference.

We also conduct the same analysis with the hosts of core-collapse SNe \citep{galb17}, and no statistically significant results are obtained.

\movetableright=-16mm
\begin{table*}
\caption{Host Galaxy Properties} \label{tab:properties}
\footnotesize
\begin{tabular}{cCcCCcCCC}
\hline\hline
FRB & $M_{\rm gas}${}$^{\rm a}$ & Ref & $M_*$ & \colhead{SFR} & Ref & 
$\mu_{\rm gas}{}^{\rm b}$ & $\tau_{\rm depl}{}^{\rm c}$ & \colhead{SFE$^{\rm d}$}\\
 & (M_\odot) & & (M_\odot) & \colhead($M_\odot$ yr$^{-1}$) & 
 & \colhead{(Gyr)} & \colhead{(Gyr$^{-1}$)} \\
\hline
20121102A & <9.1\times10^9 & 1 & (1.4\pm0.7)\times10^9 & 0.15\pm0.04 & 4 & 
65 & <61 & >0.017 \\
20180924B & (2.4\pm0.2)\times10^9 & 1 & (1.32\pm0.51)\times10^{10} & 0.88\pm0.26 & 4 &
0.18\pm0.07               &  2.8\pm0.8               & 0.36\pm0.11 \\
20190102C & <3.8\times10^8 & 1 & (3.39\pm1.02)\times10^9 & 0.86\pm0.26 & 4 &
<0.11                      & <0.45                    & >2.2 \\
20190711A & <6.7\times10^9 & 1 & (8.1\pm2.9)\times10^8 & 0.42\pm0.12 & 4 &
<8.3                      & <16                     & >0.063 \\
20200120E & 4\times10^7 & 2 & (7.2\pm1.7)\times10^{10} & 0.6\pm0.2 & 4 &
(5.6\pm1.3)\times10^{-4} & 0.067\pm0.022 & 15\pm5 \\
20200428A & 2.5\times10^9 & 3 & (6.08\pm1.14)\times10^{10} & 1.65\pm0.19 & 5 & 
0.041\pm0.008 & 1.5\pm0.2 & 0.66\pm0.08 \\
\hline
\end{tabular}
\tablerefs{
(1) This work; 
(2) \cite{brou91}; 
(3) \cite{kalb09}; 
(4) \cite{hein20a}
(5) \cite{licq15}; 
}
\tablecomments{
Limits are 3$\sigma$. \\
$^{\rm a}$ Molecular gas mass derived with the Milky Way conversion factor ($\alpha_{\rm CO} = 4.3$ $M_{\odot}$ (K km~s$^{-1}$ pc$^2$)$^{-1}$) except for 
the FRB\,20121102A host ($\alpha_{\rm CO} = 200$ $M_{\odot}$ (K km~s$^{-1}$ pc$^2$)$^{-1}$) and 
the FRB\,20190711A host ($\alpha_{\rm CO} = 60$ $M_{\odot}$ (K km~s$^{-1}$ pc$^2$)$^{-1}$). \\
$^{\rm b}$ Molecular gas mass fraction ($M_{\rm gas}/M_*$). \\
$^{\rm c}$ Gas depletion time ($M_{\rm gas}/$SFR). \\ 
$^{\rm d}$ Star-formation efficiency (SFR$/M_{\rm gas}$). 
}
\end{table*}

\section{Conclusions and Implication for Progenitors} \label{sec:conclusions}

In this study, we report molecular gas properties of six FRB hosts obtained from ALMA observations (20180924B, 20190102C, and 20190711A) and in the literature (20121102A, 20200120E, and 20200428A), allowing comparisons of a sample of FRB hosts with other galaxy populations for the first time. 
We found the diversity of molecular gas properties of the FRB hosts with wide ranges of molecular gas mass, gas depletion time, and gas mass fraction. 
Compared to other star-forming galaxy populations, the FRB\,20180924B host is molecular gas-rich, and the hosts of FRBs 20190102C and 20200120E are molecular gas-poor with a shorter gas depletion for their stellar mass and SFR.

Optical studies found various properties in FRB hosts, such as morphologies, colors, luminosities, stellar masses, SFRs, ages, or emission-line ratios in the BPT diagram, and they do not track the distribution of field galaxies \citep[e.g.,][]{hein20a, bhan22}.
Our study adds a new perspective of molecular gas to the diversity of FRB hosts. 
If FRBs originate from massive stars (similar to core-collapse SNe or LGRBs) or their remnants, an adequate quantity of molecular gas is expected for producing the progenitors. 
In contrast, less molecular gas is expected in such environments if FRBs originate from old populations (such as white dwarfs, old neutron stars, and stellar-mass black holes). 
Although we cannot constrain specific progenitor models, our findings suggest that FRBs arise from multiple progenitors or from a single progenitor that can exist in a wide range of host galaxy environments.

Currently, the limited sample size makes statistical comparisons with other galaxy environments difficult. 
Ongoing and future molecular gas observations will overcome this situation as the number of detected FRBs and their localized host galaxies rapidly increases. 
In addition, it is essential to spatially resolve the host galaxies to determine the nature of the molecular gas at the FRB positions in detail. 
This would help settle the debate over whether the progenitors of FRBs are old populations or not.

\begin{acknowledgments}
We would like to acknowledge the reviewers for valuable comments and suggestions which significantly improved the Letter. 
We thank ALMA staffs for data acquisition. 
BH is supported by JSPS KAKENHI Grant Number 19K03925. 
TH acknowledges the supports of the Ministry of Science and Technology of Taiwan through grants 110-2112-M-005-013-MY3, 110-2112-M-007-034-, and 111-2123-M-001-008-. 
T-YH is supported by the University Consortium of ALMA-Taiwan Center for ALMA Science Advancement through a grant 110-2112-M-007-034-.
This work is supported by the ALMA Japan Research Grant of NAOJ Chile Observatory (NAOJ-ALMA-268). 
\end{acknowledgments}
\begin{acknowledgments}
This paper makes use of the following ALMA data: ADS/JAO.ALMA\#2021.1.00027.S and \#2019.01450.S. ALMA is a partnership of ESO (representing its member states), NSF (USA) and NINS (Japan), together with NRC (Canada), MOST and ASIAA (Taiwan), and KASI (Republic of Korea), in cooperation with the Republic of Chile. The Joint ALMA Observatory is operated by ESO, AUI/NRAO and NAOJ.
\end{acknowledgments}

\facility{ALMA.}

\software{
CASA \citep{mcmu07}, 
{\sc astropy} \citep{the13, the18c}, 
}


\listofchanges
\end{document}